          [2006/09/06]
\documentclass[a4paper,twocolumn]{esapub2005} 
\usepackage{times}
\usepackage{natbib}
\usepackage{color}
\usepackage{graphics}
\usepackage{epsfig}

\title{Pulsation properties of the $\beta$ Cephei star SY  Equ from combined
      photometric and spectroscopic data$^{*}$}

\author{Jadwiga Daszy\'nska-Daszkiewicz}
\author{Andrzej Pigulski}
\affil{Instytut Astronomiczny, Uniwersytet Wroc{\l}awski, ul. Kopernika 11, 53-614 Wroc{\l}aw, Poland\\
E-mails: daszynska@astro.uni.wroc.pl, pigulski@astro.uni.wroc.pl}

\begin{document}


\thanks{Based on photometry obtained
      at the Bia{\l}k\'ow Observatory (University of Wroc{\l}aw) and on spectroscopy
      obtained at the David Dunlap Observatory (University of Toronto)}

\maketitle

\begin{abstract}
We present the analysis of simultaneous multicolour $uvyI_{\rm C}$
photometry and low-resolution spectroscopy for the rapidly rotating
$\beta$ Cephei star SY Equ. From the photometric time series, we confirm the
dominant pulsation frequency, $f_1$ = 6.029 d$^{-1}$, and we find an
evidence for two additional modes. In spectroscopy, the highest peak
occurs at $f_{\rm a}$ = 0.197 d$^{-1}$ or its alias 0.803 d$^{-1}$. It can be
interpreted either in terms of a binary motion or as the $g$-mode
pulsation. In addition, we reveal the pulsation mode with frequency of
about 6.029~d$^{-1}$, i.e. the same which dominates photometric variations,
and a few new candidates. For the dominant frequency we obtain mode identification
from the combined photometric and spectroscopic observations. From non-adiabatic
pulsation calculations, we show that the frequency of the dominant mode in SY Equ
is consistent with the stellar models of much lower effective
temperatures than used in many papers.
\end{abstract}

\section{Introduction}

SY Equ (HD 203664) is the early-B spectral type star, B0.5\,IIIn, classified
as the $\beta$ Cephei variable by \cite{aerts2000} from the Hipparcos photometry.
The star is one of the fastest rotators among $\beta$ Cephei pulsators, having
projected rotational velocity of about 200 km~s$^{-1}$ \cite{little1994}.
Moreover, it is a high Galactic latitude star with $b$ = $-$27.5$^{\circ}$ which locates
it more than 1~kpc out of the Galactic plane.

There are several determinations of stellar parameters for SY Equ.
\cite{little1994} derived $\log T_{\rm eff}$ = 4.447~dex, $\log g$ = 3.7~dex
from the high-resolution optical spectra and estimated a mass of 14~$M_{\odot}$
from the evolutionary tracks of \cite{mame88}. On the other hand,
using Geneva photometry, \cite{aerts2006} determined
$\log T_{\rm eff}$ = 4.47 $\pm$ 0.01 dex and $\log g$ = 3.9 $\pm$ 0.3~dex.
Much lower value of effective temperature, $\log T_{\rm eff}$ = 4.401~dex, was obtained
by \cite{keenan1982} from Str\"omgren photometry and equivalent width of the H$\epsilon$ line.
We adopt here stellar parameters derived from the International Ultraviolet Explorer (IUE)
spectra and visual photometry by \cite{niemdaszy2005}: $\log
T_{\rm eff}$ = 4.388 $\pm$ 0.035 and $\log g$ = 3.905 $\pm$ 0.1. These values
result in a mass estimate of 11.5 $\pm$ 1.5~$M_{\odot}$ if the metallicity $Z$ = 0.02 is assumed. The
metallicity in terms of [m/H], obtained from the IUE/INES data, amounts to
0.00 $\pm$ 0.21~dex, which corresponds to $Z$ = 0.02$^{+0.012}_{-0.007}$.
We prefer to rely on this determination because it uses ultraviolet spectral region,
where B-type stars emit most of their energy.

In this note we give only a brief outline of our results.
The full analysis will be published elsewhere.

\section{Observations}
\subsection{Photometry}
The new CCD photometry of SY Equ was carried out in Bia{\l}k\'ow Observatory (Wroc{\l}aw University)
during five nights between August 14 and September 3, 2004.
We used Str\"omgren $uvy$ and Cousins $I_{\rm C}$ passbands.
The obtained frames were calibrated in a standard way and then reduced using
the profile-fitting and aperture photometry of the Daophot package \cite{stetson1987}.

\subsection{Spectroscopy}

\begin{figure}
\centering
\includegraphics[width=80mm, clip]{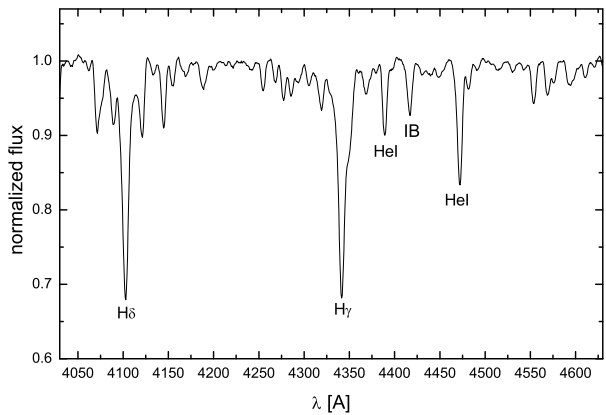}
\caption{The mean spectrum of SY Equ in the observed range obtained by averaging
         of 171 normalized individual spectra. IB stands for interstellar band.}
\label{mean_spectrum}
\end{figure}

Spectroscopic observations of SY Equ were obtained with the Cassegrain spectrograph
attached to the 1.88-m telescope of the David Dunlap Observatory (University of Toronto).
The data consist of 171 spectra taken on 12 nights between August 8 and September 1, 2004,
i.e., simultaneously with the Bia{\l}k\'ow photomety.
The spectra were sampled over the range 4000--4600~{\AA} with the 600 lines per mm grating
giving a dispersion of 0.62~{\AA} per pixel. The exposure times were 15 minutes long.
The reductions were made with standard IRAF\footnote{IRAF is distributed by the National
Optical Astronomy Observatories, which are operated by the Association of Universities for
Research in Astronomy, Inc., under cooperative agreement with the National Science Foundation.}
routines, which include cosmic ray removal, bias correction, flat-fielding,
and wavelength calibration. The spectra were moved to the heliocentric frame.
In Fig.~\ref{mean_spectrum} we show the mean spectrum of SY Equ.

\section{Frequency analysis}

\begin{figure}
\centering
\includegraphics[width=80mm, clip]{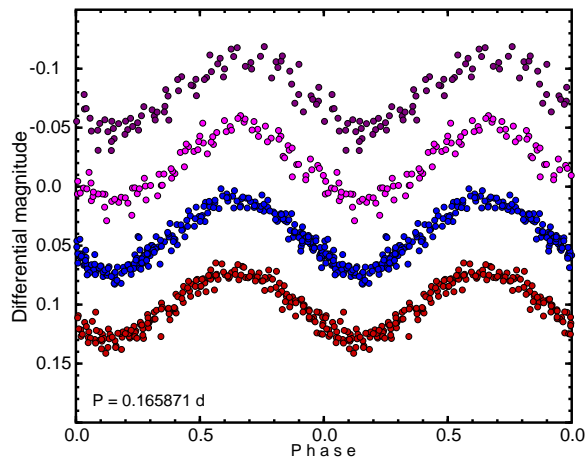}
\caption{Phase diagrams of the Bia{\l}k\'ow 2004 SY Equ light-curves folded with the period of 0.165871~d.
Going from top to bottom, data in $u$, $v$, $y$, and $I_{\rm C}$ photometric passbands are shown.}
\label{Bialkow-phot}
\end{figure}

\begin{figure}[!ht]
\centering
\includegraphics[width=80mm, clip]{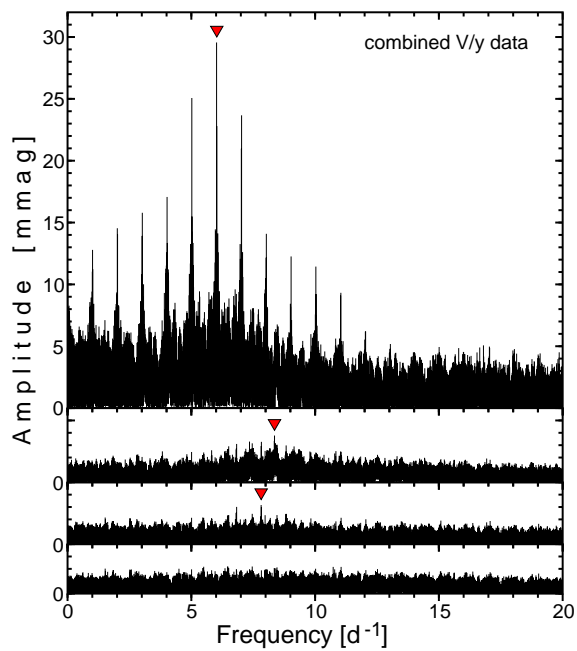}
\caption{Fourier periodogram of the of the combined V/y photometric data of SY Equ (top)
         and after consecutive steps of prewhitening (lower panels).}
\label{FP-phot}
\end{figure}

The photometric light curve of SY Equ is dominated by the mode with frequency $f$ = 6.029 d$^{-1}$.
The 2004 Bia{\l}k\'ow $vbyI_{\rm C}$ photometry folded with the period of 0.165871~d = 1/$f$
is shown in Fig.~\ref{Bialkow-phot}. In order to search for possible low-amplitude modes,
we combined three photometric datasets: Geneva $V$-band data of \cite{aerts2006},
ASAS-3 $V$-band photometry \cite{pojm2005,pigu2005} and Bia{\l}k\'ow 2004 data in
the Str\"omgren $y$ band. These data cover the time interval of almost 9 years, between 1997 and 2005.
In Fig.~\ref{FP-phot} we plotted Fourier periodograms of the combined $V$/$y$ data showing consecutive
steps of prewhitening. We see that except for the well-known dominating mode,
at least two low-amplitude modes are present.  Their frequencies are equal to 8.360 and 7.821~d$^{-1}$,
but due to the severe aliasing, their values are not certain.

For the spectroscopic data, we searched for intrinsic variations in the first four moments
of four spectral lines: H$\gamma$, H$\delta$, He\,4471, and He\,4387, labeled in Fig.~\ref{mean_spectrum}.
Moreover, we did the same for radial velocities obtained by means of the cross-correlation
function (CCF) technique using the whole observed spectrum excluding the region of the intestellar
band (IB, Fig.~\ref{mean_spectrum}). The frequencies of the terms we found are summarized in Tab.~1.
\begin{table*}
  \begin{center}
    \caption{Frequencies [in d$^{-1}$] detected with signal-to-noise (S/N) ratio greater than 4 in the first
     three moments of H$\gamma$ and H$\delta$ lines, and CCF radial velocities. The frequencies are given
     in the order of detection. Numbers in parentheses denote the r.m.s.~errors of the last digits.}\vspace{1em}
\begin{tabular}[h]{*{6}{|c}|c|} \hline
\multicolumn{3}{|c|}{H$\gamma$} &\multicolumn{3}{|c|}{H$\delta$} &    \\
\hline\hline
 $M_0$  &   $M_1$ &   $M_2$ &  $M_0$  & $M_1$ & $M_2$ & $V_{\rm rad}(CCF)$ \\
\hline\hline
 6.0318(39) & 0.1970(15) & 0.1921(27) &  0.4683(42) & 0.1970(16) & 6.0354(14) & 6.0241(12)   \\
 & 6.0267(16) & 6.0283(30) & 6.0284(37) & 6.0253(18) & 0.3532(21)  & 0.1978(19) \\
& 0.6094(21) &            &            & 4.4205(33) &             & 4.4281(28) \\
& 9.0304(28) &            &            &            &             & 0.6026(36) \\
\hline
    \end{tabular}
  \end{center}
\label{tab-freq}
\end{table*}
In all lines and the CCF radial velocity, we have found
low-frequency variation at 0.197 d$^{-1}$, which did not appear in
photometry, and the peak known already from photometry,
$\sim$6.029~d$^{-1}$. The radial velocity amplitudes of these two
frequencies are 15 and 12 km s$^{-1}$, respectively. Additionally,
we found low-amplitude changes with 4.42~d$^{-1}$ and 0.61~d$^{-1}$.
The latter could be the rotation frequency of SY Equ. In Table 1 we
give frequencies which were identified after subsequent stages of
prewhitening for three moments of two lines (H$\gamma$ and
H$\delta$) and $V_{\rm rad}$(CCF). The important results is that we
did not find the variation with frequency 2$f_1$, where $f_1$ is the
frequency of the dominant mode, in the second moments of the
analyzed four lines. This leads immediately to the conclusion that
the dominant frequency has to be a non-axisymmetric mode ($m \ne$0).

Recently \cite{aerts2006} suggested two additional modes with frequencies of 6.82902 and 4.81543~d$^{-1}$,
from the analysis of the Geneva photometry. The former is probably an alias of our mode with 7.821~d$^{-1}$,
the presence of the latter was not confirmed in our analysis.

\section{Mode identification}
\begin{figure}[!ht]
\centering
\includegraphics[width=80mm, clip]{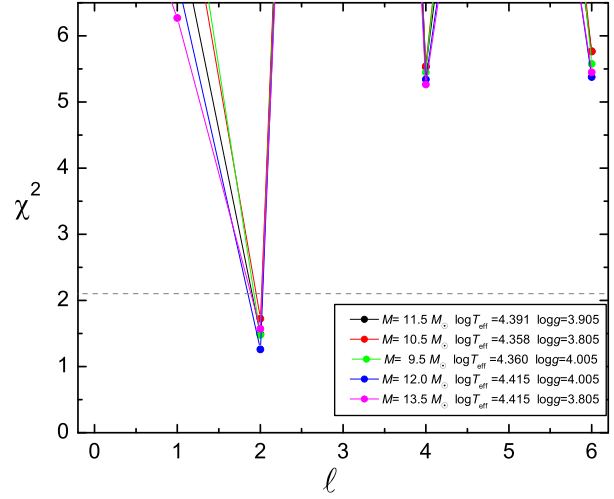}
\caption{The value of $\chi^2$ as a function of $\ell$ obtained for five models of SY Equ.
The models are located in the error box defined in Sect.~1.}
\label{l-ident}
\end{figure}

\begin{figure}[!ht]
\centering
\includegraphics[width=80mm, clip]{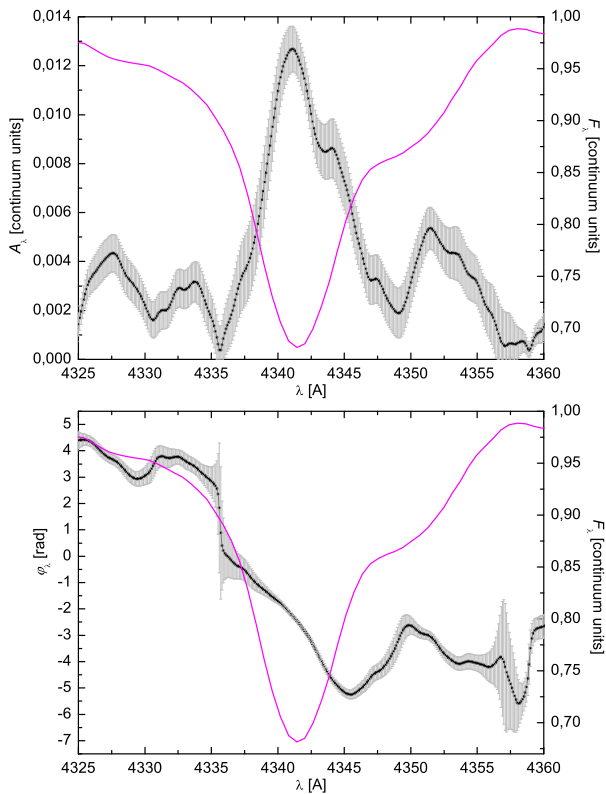}
\caption{The monochromatic amplitude (top) and phase (bottom)
diagrams for H$\gamma$. The line profile is plotted with a thin line
and the amplitudes and phases with a thick one. Errors are shown as
grey shadows. Right ordinate is given in flux units and refers to
the line profile.} \label{m-ident}
\end{figure}

We determined the amplitudes and phases of the light curves in
$uvyI_{\rm C}$ bands and of the radial velocity variations. Using
these data for the dominant frequency, we identified its spherical
harmonic degree, $\ell$, by means of the method invented by
Daszy\'nska, Dziembowski and Pamyatnykh \cite{daszy2003,daszy2005}.
The method combines photometry and spectroscopy and consists of
simultaneous determination of the $\ell$ value and the pulsation
nonadiabatic parameter $f$. The $f$ parameter describes the ratio of
the relative luminosity variation to the radial displacement of the
surface, and it gives information on subphotospheric layer. In the
case of $\delta$ Scuti stars, this new diagnostic tool probes the
efficiency of convection \cite{daszy2003}, while in the case of
$\beta$ Cep stars, it provides a stringent probe of stellar
opacities \cite{daszy2005}. From this method, we get the unambiguous
$\ell =2$ identification. In Fig.~\ref{l-ident}, we show the
$\chi^2$ value as a function of $\ell$. The horizontal line at
$\chi^2$ = 2.1 corresponds to 95\% confidence level. In the next
step, we identified the azimuthal order, $m$, from spectroscopy,
using monochromatic amplitude and phase diagrams across the line
profiles. In Fig.~\ref{m-ident} we plot these diagrams across
H$\gamma$. As one can see, the amplitude reaches a very sharp
maximum in the line center, which indicates $|m|=2$ as the most
probable identification. The sign of $m$ is obtained from the phase
changes. Because the phase decreases across the line profile with
increasing wavelengths, the mode must be prograde, $m=+2$. Thus, the
dominant frequency of SY Equ is a sectoral prograde mode with
$\ell=2$. Such mode identification excludes low inclination angles
because sectoral modes are poorly seen when the star is observed
near pole-on.

\section{Pulsation modeling}
Having identified the geometry of the main pulsation mode in terms
of $\ell$ and $m$ parameters, we may compare the observed frequency
of this mode with the theoretical predictions. For this purpose,
stellar evolutionary models were calculated using Warsaw-New Jersey
evolutionary code, assuming the OPAL opacities and the standard
solar mixture of elements. Linear nonadiabatic pulsation
calculations were performed with the use of Dziembowski's code,
which includes the rotation effects on oscillation frequencies up to
the second order in the rotational velocity. Although SY Equ rotates
very fast (at least 35\% of the break-up velocity), we can neglect
deviations from the spherical symmetry due to effects of centrifugal
force, because the ratio of the rotation frequency to pulsation
frequency is still much less than 0.5.

\begin{figure}[!ht]
\centering
\includegraphics[width=80mm, clip]{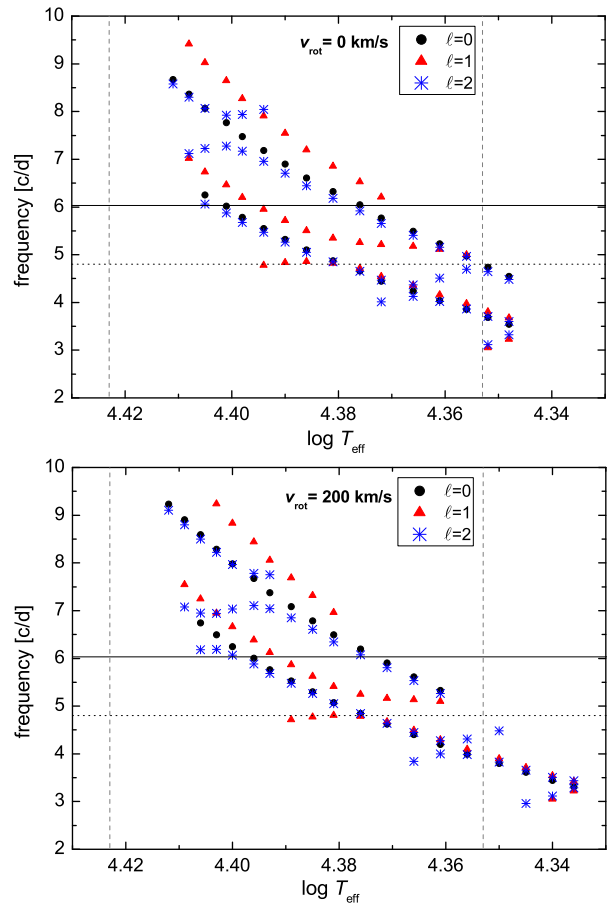}
\caption{Evolution of unstable modes with $\ell$ = 0, 1, and 2 in a sequence of
$\beta$ Cephei star models with mass of 12 $M_\odot$, assuming $Z=0.02$.
The models cover the phase from ZAMS to TAMS. The two panels show the effects of
rotational velocity (0 and 200 km\,s$^{-1}$) on the pulsation frequencies.}
\label{freq-M12}
\end{figure}

In Fig.~\ref{freq-M12}, unstable modes with $\ell$ = 0, 1, and 2 for
models with a mass of 12~$M_{\odot}$, evolving  from ZAMS to TAMS,
are shown. The initial hydrogen abundance was $X$ = 0.7 and the
heavy element abundance $Z$ = 0.02. We consider two values of the
rotational velocity, $v_{\rm rot}$= 0 and 200 km\,s$^{-1}$. The
vertical lines mark the range of effective temperatures determined
from the IUE spectra. The rotational velocity of SY Equ is equal to
at least 200 km s$^{-1}$, which gives rotational frequency of
$\sim$0.6~d$^{-1}$ assuming the stellar radius of about 6.5
$R_{\odot}$. If the frequency of 6.029 d$^{-1}$ (solid horizontal
line) is a sectoral prograde mode, we have to compare theoretical
values with the centroid of the $\ell$ = 2 mode which should be
placed at about 4.8 d$^{-1}$  (dotted horizontal line).
As one can see from Fig.~\ref{freq-M12}, for $Z$ = 0.02 the pulsation instability
begins at $\log T_{\rm eff}$ of about 4.41~dex, which is much lower
than the effective temperature of SY Equ used by some authors, e.g. \cite{aerts2006} and \cite{little1994}.
Moreover, we checked the metallicity and mass effects on the instability domain considering
several values of $Z$ and $M$. Changing $Z$ from 0.015 to 0.025 shifts the hottest unstable
mode from $\log T_{\rm eff}\approx$ 4.40 to 4.42~dex. Changing mass from 11 to 13~$M_\odot$
gives exactly the same effect on the hot edge of instability.

\section{Conclusions}

We presented the results of the frequency analysis of both the
photometric and spectroscopic time series of SY~Equ. In both datasets,
we found the dominant mode at $f_1$ = 6.029~d$^{-1}$, and several new candidate frequencies.
Moreover, there is a low-frequency variation at $f_a$ = 0.197~d$^{-1}$ detected in radial velocities,
which can be an evidence of the binary motion or the excitation of a gravity mode. For the
dominant mode we confirmed the $\ell$ = 2 identification of \cite{aerts2006} and
identified the azimuthal order, $m$ = $+$2. It means that the dominant frequency
is a sectoral prograde mode with $\ell$ =2.
However, one has to be aware that at $v_{\rm rot} \approx$ 200 km\,s$^{-1}$,
the rotational mode coupling can take place, if the frequency difference
between modes $j$ and $k$ is of the order of angular velocity of rotation,
and if the spherical harmonic indices satisfy the relation $l_j = l_k\pm$ 2, see \cite{daszy2002}.

Recently,  \cite{aerts2006} made an attempt of comparisons of
theoretical frequencies with the observed value of the dominant
mode, but the authors totally ignored the effects of rotation and
the instability condition. Our analysis showed that only for lower
effective temperatures we enter the instability domain. These
temperature values agree very well with the range determined from
the ultraviolet IUE spectra. Moreover the rotation effects on
frequencies cannot be neglected for such rapid rotator as SY Equ.

\section*{Acknowledgments}
We thank Wojtek Dziembowski and Alosha Pamyatnykh for allowing the use of the
Warsaw-New Jersey evolutionary and the non-adiabatic pulsation
codes. JDD gratefully thanks S{\l}awek Ruci\'nski for instructive
discussions and the hospitality during her stay at the DDO. We also thank
G.\,Kopacki, Z.\,Ko{\l}aczkowski and M.\,St{\c e}\'slicki for making some
photometric observations of SY Equ. The work of JDD has been
supported by the Polish MNiI Grant No. 1 P03D 021 28.

\end{document}